# Categorical Data Analysis

Dandan Chen   &   Carolyn J. Anderson
dandan3@illinois.edu   cja@illinois.edu
Department of Educational Psychology
University of Illinois Urbana-Champaign

## Abstract

Categorical data are common in educational and social science research; however, methods for its analysis are generally not covered in introductory statistics courses. This chapter overviews fundamental concepts and methods in categorical data analysis. It describes and illustrates the analysis of contingency tables given different sampling processes and distributions, estimation of probabilities, hypothesis testing, measures of associations, and tests of no association with nominal variables, as well as the test of linear association with ordinal variables. Three data sets illustrate fatal police shootings in the United States, clinical trials of the Moderna vaccine, and responses to General Social Survey questions.

*Word count*: 4639

## 1. Introduction

Valid inference requires proper methods to handle discrete data. Categorical data consist of counts or discretely measured variables. Categorical data typically consist of counts or frequencies within a category. Examples of categorical variables include race or ethnicity, responses to a multiple-choice question, status of an online purchase, students' letter grades, gender, responses to survey items, graduation status, and whether an individual receives a vaccination. The methods described in this chapter are for analyses where a categorical variable is the response, outcome, or dependent variable; predictor or explanatory variables are either categorical or continuous. The levels of categorical variables can be either nominal (e.g., type of high school attended) or ordinal (e.g., letter grade).

Whether a variable is nominal or ordinal depends on context. For example, whether you received a vaccine seems to be inherently nominal. Either you have or have not received a vaccine. However, if the question "Have you received a COVID-19 vaccine?" is an item on a scale developed to measure a person's perceived susceptibility or the consequences of getting sick, the "yes" and "no" categories are ordered concerning what the scale is measuring.

In this chapter, we use examples of two categorical variables. Fundamentals here include contingency tables, sampling processes, sampling distributions, estimation of probabilities with hypothesis testing, measures of association, and testing for relationships using statistics based on different distributions.

## 2. Notations

Throughout this chapter, notations for categorical variables are in Roman letters, and parameters of models and distributions are represented by lower-case Greek letters. For instance, the data can be presented as a two-way contingency table with two categorical variables. In this situation, $X$ and $Y$ denote categorical variables; a cross-classification of these two variables is shown in a contingency table, where $n_{ij}$ is a count or frequency within the cell $(i, j)$ (i.e., row $i$ and column $j$), $n_{i+}$ and $n_{+j}$ are the row and marginal column totals, and $n$ is the total sample size. When there are more than two rows or columns, the numbers of rows and columns are denoted by $I$ and $J$, respectively. $\mu$ represents the mean of a variable (X or Y), $\pi$ denotes a probability, $p$ is a sample proportion. For example, $\mu_{ij}$, $\pi_{ij}$ and $p_{ij} = n_{ij}/n$ represent the mean, probability, and sample proportion for cell $(i, j)$, respectively. $df$ is the short form for degrees of freedom.

3. **Sampling Processes**

The data about fatal shootings from on-duty police officers in the United States from January 2015 to December 2020, compiled by the Washington Post, were cross-classified in Table 1 for two variables: gender (female, male) and race (non-white, white). We dropped three incidents that involved young children. The values within each cell are the frequencies of cases for each combination of gender and race, and the numbers in the margins are totals by row and by column.

**Table 1.** Counts of fatal police shootings cross-classified in a two-way table by race and gender.

|  | Woman | Man |  |
|---|---|---|---|
| Non-White | 98 | 2,892 | 2,990 |
| White | 155 | 2,552 | 2,707 |
|  | 253 | 5,444 | 5,697 |

Incidents in Table 1 are assumed to be independent of each other, and each incident falls into only one contingency table cell. In the theory of population sampling, the data in Table 1 reflect the *Poisson sampling*, in which each population unit has a certain probability of being included in the sample data (Särndal et al., 1992). In this case, the marginal frequencies and the total number of observations are not bounded or fixed by design; therefore, cell proportions and the row and column proportions reflect actual probability distributions in the population, which can differ.

Table 2 shows an example of independent *Bernoulli sampling*, a special Poisson sampling case. The probability of being sampled into a category for each subject is the same in this sampling process. Specifically, this probability remains .5 because there are only two categories for the corresponding variable. Data in Table 2 offer an example. They come from a phase-3 clinical trial of the Moderna vaccine for COVID-19. In this trial, 15,210 subjects were randomly assigned to the placebo and the vaccine groups (Baden et al., 2020), and they were later classified as being "symptomatic" or "asymptomatic." In Table 2, the row margins are fixed by design.

**Table 2.** A 2 × 2 table of counts of treatment group by COVID-19 infection.

|  | Asymptomatic | Symptomatic |  |
|---|---|---|---|
| Placebo | 15,025 | 185 | 15,210 |
| Vaccine | 15,199 | 11 | 15,210 |
|  | 30,224 | 196 | 30,420 |

4. **Sampling Distributions**

When binomial sampling is used, as shown in Table 2, the row distributions are independent, and row margins follow a *binomial distribution*. Specifically, the *binomial distribution* is helpful for modeling counts or proportions of a variable (or event) that has two categories (or outcomes), such as students' success or failure in a college course and getting head or tail when flipping a coin. The binomial distribution assumes that (1) there is a fixed total of independent trials for this event, in which each trial has two outcomes (known as a *Bernoulli trial*), and that (2) the probability of the targeted outcome (e.g., "success") is the same for each trial. A Bernoulli trial with the quality of (2) is also known as an *identical trial*.

The mean and variance of the probability of the outcome $Y$ in a fixed number of Bernoulli trials are

$$E(Y) = n\pi, VAR(Y) = n\pi(1-\pi), \qquad (1)$$

where $n$ denotes the total number of trials and $\pi$ denotes the probability of the occurrence of the targeted in one trial. The probability of having the targeted outcome occurring in $y$ trials follows the binomial probability mass function:

$$P(y) = \frac{n!}{y!(n-y)!}\pi^y(1-\pi)^{n-y}, y = 0, 1, 2, \dots, n \qquad (2)$$

For example, if the probability of selecting a student enrolled in the National School Lunch Program in a school district is .2, then the probability of having seven students who are enrolled in this program, out of a random sample of 10 students in this school district, is $P(7) = \frac{10!}{7!(10-7)!} \cdot .2^7(1-.2)^{10-7} = .0008$, which is .08%. It is worth noting that the sample size matters here to ensure accuracy in making statistical inferences: The fixed total $n$ should be at least 10 when $\pi$ equals .5 to make the binomial distribution symmetric. This requirement is necessary to meet normality, a fundamental assumption undergirding frequentist statistical modeling. The required minimum gets larger when $\pi$ gets closer to 0 or 1 to achieve this symmetric, bell-shaped distribution. In this case, we say the sample size is "large enough" and see the binomial distribution as a normal one.

In a more generalized case, if the "response" or outcome variable (i.e., symptomatic) has three or more categories (e.g., none, mild, severe), we would get contingency tables with three or more columns, and each row would follow a *multinomial distribution*. The multinomial distribution assumes that (1) there is a fixed total of independent trials for this event, in which each trial has more than two outcomes, and that (2) the probability of each outcome (e.g., none, mild, severe) is the same for each trial. When the probability of each outcome is denoted by $\pi_j$ (such that we have $\pi_1, \pi_2, \dots, \pi_J$), we must have $\sum_j \pi_j = 1$. Given the fixed total of identical trials, $n$, we must have the numbers of occurrences for each outcome, denoted by $y_j$ (such that we have $y_1, y_2, \dots, y_J$), meet $\sum_j y_j = n$. Hence, the probability of having $y_1$ trials with the first outcome, $y_2$ trials with the second outcome, …, and $y_j$ trials with the last outcome follow the multinomial probability mass function below:

$$P(y_1, y_2, \dots, y_j) = \left(\frac{n!}{y_1! y_2! \dots y_j!}\right)\pi_1^{y_1}\pi_2^{y_2} \dots \pi_j^{y_j}, y_j = 0, 1, 2, \dots, n \qquad (3)$$

Evidently, the binomial distribution described above is a special case of the multinomial distribution in which *j*=2.

The *Poisson distribution* is typically used for counts or proportions of an event occurring randomly over time or space. It helps analyze the categorical data, such as the number of students that completed an activity online in an hour of a day, and the number of people who arrive at a university gym within 5 minutes during Day 1 of a new semester, as shown in Table 3. This distribution assumes the following:

**Table 3.** Example: Number of people who arrive at a university gym during Day 1.

| Time | Observed frequency |
|---|---|
| 8:00 am – 8:05 am | 2 |
| 8:00 am – 8:05 am | 11 |
| 8:00 am – 8:05 am | 32 |
| … | … |

(1) The numbers of people arriving at the university gym in non-overlapping time intervals are independent, (2) the probability of precisely one person arriving at a university gym, $\pi$, is proportional to the length of the time interval, and (3) the probability of two or more people arriving at the university gym in a sufficiently short interval is essentially 0. This Poisson probability distribution is unimodal and skews to the right. Its mean and the variance (of the probability of occurring) for the outcome $Y$ are equal to $\pi$:

$$E(Y) = VAR(Y) = \pi. \tag{4}$$

For the outcome $Y$ occurring in $y$ trials, the Poisson probability mass function is

$$P(y) = \frac{e^{-\mu}\mu^y}{y!}, y = 0, 1, 2, \ldots, n. \tag{5}$$

As $\pi$ increases, the Poisson distribution becomes more symmetrical and closer to a bell shape.

## 5. Types of Probability

For two-way tables, we distinguish between three types of probability: joint probability, marginal probability, and conditional probability. The *joint probability* of $X$ and $Y$ is the probability that an observation falls into cell $(i, j)$, a combination of categories from two variables. For example, given the fatal police shooting in Table 1, we can calculate maximum likelihood estimates for the joint probabilities, as shown in Table 4. For example, the estimated joint probability that the victim was a non-white woman equals $\hat{\pi}_{11}$=.017.

**Table 4**. Maximum likelihood estimates of joint and marginal probabilities of fatal police shootings.

|  | Woman | Man |  |
|---|---|---|---|
| Non-White | 98/5697 = .017 | 2892/5697 = .508 | 2990/5697 = .525 |
| White | 155/5697 =.027 | 2552/5697 = .448 | 2707/5697 = .475 |
|  | 253/5697= .044 | 5444/5697= .956 | 1.000 |

In some situations, our focus may be on the row or column margin, investigating the *marginal probability*. The marginal probabilities equal $\pi_{i+} = \pi_{i1} + \pi_{i2} + \cdots + \pi_{iJ}$ for row margins, and $\pi_{+j} = \pi_{1j} + \pi_{2j} + \cdots + \pi_{Ij}$ for column margins. For instance, using the police shooting data in Table 1, we can obtain maximum likelihood estimates of marginal probabilities of victims being non-white or white, or of marginal probabilities of victims being woman or man. They are shown in the row and column margins of Table 3, respectively. This marginal probability of victims being non-white is $\hat{\pi}_{1+}$ = .525, and this marginal probability of victims being white is $\hat{\pi}_{2+}$ = .475.

The *conditional probability* refers to the probability of an outcome conditioning on the probability of another outcome. For the two-way contingency tables, the conditional probability of the category $j$ of one variable given the category $i$ of another variable equals

$$\pi_{j|i} = \pi_{ij}/\pi_{i+} .\qquad(6)$$

Given the vaccine data in Table 2, we can calculate conditional probabilities for each row and column, shown in Table 5. When the conditional probabilities for the two rows are approximately the same for a column (i.e., either the asymptomatic or symptomatic group), we may conclude that the vaccine is not effective for the asymptomatic/symptomatic group.

**Table 5**. Maximum likelihood estimates of conditional probabilities from COVID-19 vaccine data.

|  | Asymptomatic | Symptomatic |  |
|---|---|---|---|
| Placebo | 15025/15210=.9878 | 185/15210=.0122 | 1.00 |
| Vaccine | 15199/15210=.9993 | 11/15210=.0007 | 1.00 |
|  | 30224/30420=.9936 | 196/30420=.0064 | 1.00 |

## 6. Estimation of the Unknown Probability

### 6.1. Estimation Method: Maximum Likelihood Estimation

As hinted, our probability estimates presented in Tables 4-5 are *maximum likelihood estimates*. In the practice of categorical data analysis, the probability parameter $\pi$ in the binomial, multinomial, or Poisson distribution is often unknown. In this case, we need to *infer* the value of the unknown based on the knowns like $n$ and $y$, which researchers often have control of in a research design. In an inferential method, a (point) *estimate* is a value derived from some observed data. It is an inference of the *estimator*, which can be seen as a random variable that follows a specific sampling distribution. In the most popular inferential method, *maximum likelihood estimation* method, the *estimate* is a value of the e*stimator* following a *likelihood function*. The estimate maximizing this function is known as the *maximum likelihood estimate*, and it can be interpreted as the *most likely* value of the unknown parameter given the values of known parameters.

The maximum likelihood estimation method is popular because of an important property: Its estimator of probability (e.g., $\pi$) can be approximated by a sample proportion (e.g., $\hat{p}$), which follows a normal distribution when the sample size is large (i.e., the *Law of Large Numbers*, which claims that a proportion in a sample, $\hat{p}$, gets closer to the proportion in the population, $p$, as the sample size increases). This normality minimizes standard errors in parameter estimation. In other words, the maximum likelihood estimators have good "large-sample behaviors." Other estimators tend to have relatively large standard errors because they usually do not follow a normal distribution. The mean and the variance of the sampling distribution of the maximum likelihood estimator $\pi$ are the same as that of the binomial distribution for $p$:

$$E(\hat{\pi}) = E(\hat{p}) = \pi, VAR(\hat{\pi}) = VAR(\hat{p}) = \frac{\pi(1-\pi)}{n}.\qquad(7)$$

The likelihood function is mathematically equivalent to the joint probability mass function evaluated at the observed values (Hogg et al., 2015). For $n$ trials that follow the binomial distribution, when the targeted outcome shows up $y_1, y_2, \ldots, y_n$ times, this joint probability mass function is

$$P(y_1, y_2, \ldots, y_n) = \prod_{i=1}^{n} \pi^{y_i}(1-\pi)^{n-y_i} = \pi^{\Sigma y_i}(1-\pi)^{n-\Sigma y_i}. \tag{8}$$

When treated as a function of $\pi$, the above function is known as the likelihood function $L(\pi)$:

$$L(\pi) = f(y_1; \pi)f(y_2; \pi) \ldots f(y_n; \pi) = \pi^{\Sigma y_i}(1-\pi)^{n-\Sigma y_i}. \tag{9}$$

When solving this function, we can obtain the maximum likelihood estimate of the unknown parameter $\pi$, denoted by $\hat{\pi}$. This $\hat{\pi}$ equals the observed proportion $\hat{p}$ of an event in a sample, expressed as

$$\hat{\pi} = \hat{p} = \frac{y}{n}. \tag{10}$$

### 6.2. Hypothesis Testing

The maximum likelihood estimate of the unknown parameter in a population, as described above, is based on a sample. One sample is generally not a perfectly accurate reflection of the whole population. For example, while we know the probability of obtaining a head when flipping a coin is $\pi = .5$ (i.e., the population parameter), we may get three heads when flipping a coin ten times (i.e., a sample) instead of five heads. This sample statistic about the probability of getting a head is $\hat{\pi} = 3/10 = .3$, which is not .5. The difference between a sample statistic and the population parameter is called *sampling error* or *margin of error*. In this example about flipping a coin, the sampling error is .2 (i.e., from .5 to .3).

Considering the existence of the sampling error, *hypothesis testing* is commonly used as a decision-making process to draw inferences about the population based on the sample data. It is about rejecting one of two *hypotheses* related to a research question: the *null hypothesis* and the *alternative hypothesis* (also known as the scientific/research hypothesis). In the prior example about flipping a coin, we can set two competing hypotheses about whether the sample statistic is close enough to the population parameter, corresponding to a two-sided hypothesis test. The null hypothesis is

$$H_0: \pi = .5. \tag{11}$$

The alternative hypothesis is

$$H_1: \pi \neq .5. \tag{12}$$

In hypothesis testing, *p-value* is the probability of observing the data at least as extreme as the observed sample, given that the $H_0$ is true. The *alpha level*, also known as *level of significance* or *Type-I error rate*, is the probability of rejecting the null hypothesis when the null is true; it can be seen as the acceptable error level.

To examine whether the sample statistic is close enough to the population parameter, we check whether the difference between the two is *statistically significant*. If yes, we shall reject the null hypothesis and go with the alternative hypothesis, concluding that the sample statistic is not close to the population parameter. By "statistically significant," we mean the corresponding *p*-value is *lower* than the alpha level. How can we obtain the *p*-value then? Researchers use the *standard normal distribution* for converting the difference between the sample statistic and the population parameter into a $z$ score and then into a *p*-value. In the previous example about flipping a coin, the standard error of the population parameter $\pi=.5$ is calculatd as follows:

$$SE = \sqrt{VAR(\hat{\pi})} = \sqrt{\frac{\pi(1-\pi)}{n}} = \sqrt{\frac{.5(1-.5)}{10}}. \quad (13)$$

With this standard error, the difference between the sample statistic and the population parameter can be converted into a $z$ score:

$$z = \frac{\hat{\pi}-\pi}{SE} = \frac{.3-.5}{\sqrt{\frac{.5(1-.5)}{10}}} = -1.265, \quad (14)$$

which positions the sample statistic in a standard normal distribution, with which we can find the $p$-value is .898 for $z$= -1.265. When setting the alpha level at .05, which is common in educational research, we see the $p$-value .898 is not statistically significant, and so we fail to reject the null hypothesis and conclude that the sample statistic .3 is close enough to the population parameter .5.

As an alternative to the $p$-value, a *confidence interval* is a range of values centered around the sample statistic for hypothesis testing. *Confidence level*, $(1 - \alpha) \times 100\%$, defines the width of this range. It is the probability that this range contains the actual value of the population parameter, associated with the alpha level. For example, when the alpha level is .05, the confidence level is 95%. The following formula is used to construct this confidence interval for the population parameter $\pi$:

$$CI = \hat{\pi} \pm z_{\alpha/2}(\widehat{SE}), \widehat{SE} = \sqrt{\frac{\hat{\pi}(1-\hat{\pi})}{n}}, \quad (15)$$

where $\hat{\pi}$ is the sample statistic. If the confidence interval contains 0, we fail to reject the null hypothesis and conclude that the statistic $\hat{\pi}$ is not statistically significant; if not, we reject the null hypothesis. Using the prior example, we get $\widehat{SE} = \sqrt{.3(1-.3)/10}$ = .145, $z_{\alpha/2} = z_{.025}$ = 1.960, and so the confidence interval is $CI = .3 \pm 1.960 \times .145$ = (.02, .58). Because this interval does not contain 0, we reject the null hypothesis and conclude that the estimate .3 is statistically significant at the .05 level.

Broadly speaking, there are three pillars in the literature about hypothesis testing, and all of them can use a likelihood function to infer the population parameters. They are namely the likelihood-ratio test, the Wald test, and Rao's score test. All three tests are asymptotically equivalent under the null hypothesis. In other words, if the sample size is very large, we would not expect to see a noticeable difference among them. However, if the sample size is small to moderate, the values of the three statistics will diverge. The Wald test is the least reliable when sample size is small, and the likelihood ratio test and the score test are preferred in this case.

What is specified above is *Rao's score test*. In the *likelihood-ratio test*, a ratio of two maximum likelihood estimates is used as the test statistic, called the *likelihood-ratio test statistic*. It is

$$\chi^2 = -2\ln\left(\frac{l_0}{l_1}\right). \quad (16)$$

In this test statistic, $l_0$ is the maximum likelihood estimate of the parameter when $H_0$ is assumed to be true, and $l_1$ is the maximum likelihood estimate over all possible parameter values. In this case, $l_1$ is at least as large as $l_0$. This test statistic takes the natural log transform with the coefficient 2, as this conversion makes the statistic approximate a *chi-square* statistic, easier for computation. The corresponding $p$-value of this test statistic is the right-tailed probability of the chi-square distribution with $df$=1.

In the *Wald test*, we evaluate the unrestricted standard error of the estimate of the parameter $\hat{\pi}$, $SE$, by substituting the maximum likelihood estimate for the unknown parameter in the formula of the true standard error (e.g., $SE = \sqrt{\hat{\pi}(1-\hat{\pi})/n}$, because $\pi$ follows the binomial distribution). The confidence interval in the Wald test using this $SE$ is known as the *Wald confidence interval*. The test statistic in the Wald test is $z^2$, which takes the form in Equation (14), with the standard error being evaluated via the maximum likelihood estimation. It approximately follows a chi-square distribution with $df=1$. The *p*-value in this Wald test is the right-tailed probability for $z^2$ in this chi-square distribution. In this test, the distributions for hypothesis testing are often *normal distributions* or *chi-square distributions* when the sample size is large enough.

## 7. Measures of Association

The two most common measures of association for categorical variables are the odds ratios and Pearson's correlation coefficient. In this chapter, we primarily use the odds ratio as a measure and definition of association.

### 7.1. Odds ratio

*Odds* are the ratio of two conditional probabilities. For example, the odds of category $i$ given column category $j$ equal

$$odds = \pi_{i|j}/(1 - \pi_{i|j}). \tag{17}$$

Using the data in Table 1, the observed odds that a male victim was non-white (versus white) equals

$$\frac{\hat{\pi}_{non-white|man}}{\hat{\pi}_{white|man}} = \frac{p_{12}/p_{+2}}{p_{22}/p_{+2}} = \frac{n_{12}}{n_{22}} = 2892/2552 = 1.133. \tag{18}$$

The probability of the victim being a non-white man is 1.133 times the probability of the victim being a white man. That is, non-white males were more likely to be shot than white males. For women, the odds equal 98/155= .632.

The raio of odds, known as *odds ratio*, ranges from 0 to positive infinity. When the odds ratio is 1, it indicates the odds are equal, and so no relationship exists between the two variables (e.g., race and gender, as the row and column variables). Let $\theta_{i*,jj*}$ represent the odds ratio of two row categories $i$ and $i^*$ and two column categories $j$ and $j^*$. Then the odds ratio equals $\hat{\theta}_{12,12}$= 1.133/.632 = 1.792 in our example. It says that the *odds* that a victim was a non-white for men is 1.792 times the odds for women. It is worth noting that the odds ratio refers to ratios of odds and not to probabilities, as it would be wrong to state that "the probability of the victim being non-white for men is 1.792 times the probability for women."

Odds ratios can be computed when we calculate a "cross-product ratio" with joint probabilities:

$$\theta_{ii*} = (\pi_{ij}\,\pi_{i*j*})/(\pi_{i*j}\,\pi_{ij*}). \tag{19}$$

If we switch the odds in the numerator and denominator, we get .632/1.133 = 1/1.792= .558, and the interpretation would "flip": The odds that a victim was a woman for non-white victims is .558 times that for white victims.

### 7.2. Pearson Correlation Coefficient

The *Pearson correlation coefficient* measures the linear relationship between pairs of variables, which requires numerical codes for categories of each variable. For example, when two survey question items contain the unique responses "strongly disagree," "disagree," "agree," and "strongly agree," we use integers 1, 2, 3, and 4 to code these responses and denote these responses with $u_s$ and $v_s$ respectively for two question items, where $u$ and $v$ denote the items and $s = 1, 2, 3, …, n$, indexing individual respondents. Thus, if Respondent 1 chooses "disagree" on one item, the response is $u_1 = 2$; if this respondent chooses "agree" on the other item, the response is $v_1 = 3$. Given this coded data, we can set a contingency table with $I$ rows and $J$ columns corresponding to unique responses to the two items for calculation. Alternatively, we can organize the data in the respondent by item format, displaying all individual respondents' responses in two columns corresponding to the two items. In the latter case, the formula for computing correlation is

$$r_{XY} = s_{XY}/(s_X s_Y) = \frac{\sum_{s=1}^{n}(u_s-\bar{u})(v_s-\bar{v})}{\sqrt{\sum_{s=1}^{n}(u_s-\bar{u})^2}\sqrt{\sum_{s=1}^{n}(v_s-\bar{v})^2}}. \tag{20}$$

where $s_{XY}$ denotes the covariance between the two items, $s_X$ and $s_Y$ denote the standard deviation of the two items, respectively, $\bar{u}$ and $\bar{v}$ denote the means for the two items. For dichotomous variables, (20) is referred to as the *phi coefficient*. The choice of codes makes a difference for variables with more than two categories; we may choose the codes leading to the largest possible correlation, which can be found from correspondence analysis. For the police shooting data in Table 1, the correlation between gender and race is $r = -.059$.

## 8. Tests of No Association

We discuss tests for three possible null hypotheses: independence, homogeneous distributions, and linear relationship. The tests differ regarding the hypothesis being tested and thus the conclusions. After presenting the test statistics, we give examples of tests for nominal and ordinal variables.

### 8.1. Chi-square Test Statistics

Two common tests statistics have sampling distributions that are approximately chi-square for large sample sizes, requiring five or more counts per cell to ensure the shape of the distribution is chi-square. The two statistics are *Pearson's chi-square statistic*,

$$X^2 = \sum_i \sum_j \frac{(n_{ij}-\hat{\mu}_{ij})^2}{\mu_{ij}}, \tag{24}$$

and the *likelihood ratio chi-square statistic* or *deviance*,

$$G^2 = 2\sum_i \sum_j n_{ij} \log\left(\frac{n_{ij}}{\hat{\mu}_{ij}}\right), \tag{25}$$

where $\hat{\mu}_{ij}$ is the estimated expected frequency in cell $(i,j)$. Note that Pearson's chi-square statistic $X^2$ is an integral part of Pearson's correlation: $|r_{XY}| = \sqrt{X^2/n}$.

Both $X^2$ and $G^2$ equal 0 when $n_{ij} = \hat{\mu}_{ij}$ for all the cells (i.e., the observed frequencies equal the estimated expected frequencies). For testing independence and homogeneous association, specified below, $X^2$ and $G^2$ have the same value for the estimated expected frequency $\hat{\mu}_{ij} = n_{+j} n_{i+}/n$. That said, because the logic for arriving at $\hat{\mu}_{ij}$ differs in the sampling design, these two test statistics can lead to with different conclusions.

### 8.2. Independence

When the margins of a table are not fixed, we may run a test of independence to check if there is a relationship between variables. The null hypothesis of independence is $H_o: \pi_{ij} = \pi_{i+} \pi_{+j}$ for all $i$ and $j$, and the alternative hypothesis is $H_a: \pi_{ij} \neq \pi_{i+}\pi_{+j}$ for at least some combination of $i$ and $j$. Estimated expected frequencies equal $\hat{\mu}_{ij} = n_{i+}n_{+j}/n$. If the null hypothesis is true (and cell counts >5), then (24) and (25) have approximately chi-square distributed with $df = (I-1)(J-1)$.

For Table 1, testing whether victims' gender and race are independent yields $X^2$ =20.068 and $G^2$=20.137 with $df$=(2-1)(2-1)=1 and the $p$-values <.01. The data leads us to the conclusion that gender and race are related. The strength and nature of the dependence, as reflected by the odds ratio, are given in (18).

### 8.3. Homogeneous Distributions

When the data follow independent binomial or multinomial distributions, we may be interested in whether the conditional distributions of the response variable are the same (i.e., homogeneous). The null hypothesis is $H_o: \pi_{j|i} = \pi_{+j}$ or in terms of frequencies, $H_{o:} \mu_{j|i} = \pi_{+j} n_{i+}$.

Using the clinical trial data in Table 2, we test whether the distribution of participants receiving the placebo is the same as that of participants receiving the vaccine and get $X^2$=155.47, $G^2$ =187.88 with $df$=1 and the $p$-values < .01. These values align with what we find with the odds ratio, which shows a discrepancy between the placebo group and the vaccine group: The odds of people receiving the vaccine are 17.01 times the odds of people receiving the placebo for the asymptomatic group.

### 8.4. Test of Linear Association

For ordinal variables, there are more powerful association tests that use the order of categories. The *Mantel–Haenszel statistic* is

$$M^2 = (n_{++} - 1)r^2, \qquad (26)$$

where $r$ is Pearson's correlation coefficient defined in (3). The hypotheses are $H_o: \rho = 0$ and $H_a: \rho \neq 0$. $M^2$ follows a chi-square distribution with one degree of freedom. Table 6 is a cross-classification of responses to two items of the *General Social Survey 2018*. The two items are "In general, would you say your quality of life is ..." and "In general, what would you rate your physical health?". The possible responses are "Excellent," "Very good," "Good," "Fair," and "Poor." To test whether physical health and quality of life are related, we use equally spaced integers from 1 to 5 for coding response categories, and obtain $r = .599$, $M^2 = 834.937$, and $p$-value < .01. The data support the conclusion of a linear relationship.

**Table 6**. Ordinal data for two survey questions in *General Social Survey 2018*.

| Quality of life | Physical Health | | | | | |
|---|---|---|---|---|---|---|
| | Excellent | Very Good | Good | Fair | Poor | Total |
| Excellent | 221 | 160 | 66 | 29 | 2 | 478 |
| Very good | 120 | 410 | 328 | 81 | 11 | 950 |
| Good | 29 | 71 | 341 | 172 | 27 | 640 |
| Fair | 7 | 5 | 40 | 138 | 34 | 224 |
| Poor | 1 | 1 | 2 | 11 | 22 | 37 |
| Total | 378 | 647 | 777 | 431 | 96 | 2329 |

Generally speaking, to test a relationship, we needed $(I-1)(J-1)$ non-redundant odds ratios to characterize the association between two variables completely. Given our example, this number equals (5−1)(5−1)=16. Making use of the ordering categories, we can summarize the association using one single statistic, $r$ or $M^2$, which is more powerful than $X^2$ and $G^2$ in (24) and (25).

## 9. Concluding Remarks

This chapter scratches the surface of categorical data analysis, part of which overlaps with another chapter in this book about generalized linear models. For comprehensive coverage, a good starting place is Agresti (2019). Most statistical software programs have the functionality to analyze categorical data, including RStudio with R packages named "binom," "Epi," and "MASS."